\documentclass[12pt]{article}
\usepackage[sumlimits,intlimits,namelimits]{amsmath}
\usepackage{amssymb}
\usepackage[english]{babel}
\usepackage{bbm}
\usepackage{parskip}
\global\parskip 6pt

\newcommand{\de}{\delta}
\newcommand{\pa}{\partial}
\newcommand{\e}{\epsilon}

\newcommand{\be}{\begin{eqnarray}}
\newcommand{\ee}{\end{eqnarray}}
\newcommand{\non}{\nonumber}

\def\>{\rangle}
\def\<{\langle}

\begin{document}
\begin{titlepage}
\vspace{5mm}

\begin{center}
{\Large \bf 
Conformal Invariance for a Class of Scale Invariant  Theories in Four Dimensions}
\\ 
\end{center}

\vspace{25mm}
\begin{center}

{\bf
Ivo Sachs} \\

\vspace{5mm}

\footnotesize{
{\it Arnold Sommerfeld Center for Theoretical Physics, Ludwig-Maximilians University\\
Theresienstr. 37, D-80333 M\"unchen, Germany}}

\end{center}
\vspace{25mm}
\begin{abstract}
\baselineskip=14pt
For a four dimensional, unitary, diffeomorphism- and  scale invariant  quantum field theory with 
a well defined  scale current we argue that scale invariance implies conformal invariance. The proof relies on the Wess-Zumino consistency condition for the Weyl anomaly. \end{abstract}

\vfill

\end{titlepage}

There is a long standing issue in quantum field theory concerning the problem of specifying the precise condition under which a scale-invariant field theory is also invariant under the bigger group of conformal transformations  \cite{Coleman:1970je,Polchinski:1987dy}. In two space-time dimensions this has been settled long time ago by Zomolodchikov \cite{Zamolodchikov:1986gt} who showed that a unitary and scale-invariant action is necessarily conformally invariant. In four space-time dimensions this issue must be  more subtle since unitary, scale-invariant but not conformally invariant Lagrangians are known. For instance, a free two-form potential in 4 dimensions is scale invariant but not conformally invariant.  In spite of a number of non-trivial results (see \cite{Coleman:1970je}-\cite{Dymarsky:2015jia}  for a partial list of references) the issue has not been completely settled in higher dimensions. The purpose of this letter is to provide a proof for the set of unitary field theories that have the additional property of being diffeomorphism invariant, that is, they can be minimally coupled to gravity, and that their scale current is well defined. Every scale invariant theory has a (possibly vanishing) virial current  \cite{Coleman:1970je} but, although the dilatation charge is well defined, the invariance of the virial current under possible gauge symmetries of the theory is not guaranteed. In particular, this is the case for the free two-form potential mentioned above. To our knowledge, all presently known examples of unitary, scale invariant theories that are not conformal invariant are of this type. On the other hand gauge invariance of the virial current is crucial in our approach which is based on gauging the scale symmetry by introducing an extra Weyl-gauge field \cite{Iorio:1996ad}.  Concerning  diffeomorphism invariance, this is a technical assumption that  plays an important role since it allows us to equivalently represent conformal transformations as a restricted class of Weyl transformations.  
More precisely,  we consider a  translation- and scale invariant action, $S[\Phi]$, ie. invariant under 
\be
{}'x^\mu = e^{\omega}x^\mu\qquad\hbox{and}\qquad {}'\Phi({}'x) = e^{-d_\Phi\omega}\Phi(x),\label{st}
\ee
where $\omega$ is a constant and $d_\Phi$ is the canonical dimension of $\Phi$ which need not be a scalar field. The canonical Noether current associated with scale invariance is given by
\be
j^\mu =x^\nu T_{\mu\nu} +V_\mu\label{c}\;,
\ee
where $T_{\mu\nu}$ is the Belinfante-Rosenfeld tensor and  $V_\mu$ is the virial current \cite{Coleman:1970je}.  The first term in  (\ref{c}) implements the coordinate transformation  $x^\mu \to e^{\omega}x^\mu$ on the fields while \cite{Coleman:1970je,Iorio:1996ad}
\be
V_\mu=\pi_\Phi^\nu\Lambda_{\nu\mu}\Phi\,,\qquad \Lambda_{\nu\mu}=d_\Phi g_{\mu\nu}+2\Sigma_{\mu\nu}\label{vs}
\ee
implements the conformal rescaling, $\Phi \to e^{\Delta_\Phi\omega}\Phi$. Here, $\Sigma_{\mu\nu}$ is a generator of Lorentz transformations whose role is to subtract  the rank, $r$ of the tensor field $\Phi$, from its canonical dimension since the  Belinfante tensor  $T_{\mu\nu}$, by construction, already takes  into account the transformation of the Lorentz indices of $\Phi$. Therefore, the conformal weight is given by $\Delta_\Phi=d_\Phi-r$. 

If a scale invariant action can be coupled covariantly to the metric tensor $g_{\mu\nu}$ then, the action is equivalently invariant under constant Weyl transformations ( $g_{\mu\nu}\to e^{2\omega}g_{\mu\nu}$, $\Phi \to e^{-\Delta_\Phi\omega}\Phi$, $\omega= const.$) as follows from the composition of the scale transformation (\ref{st}) with the diffeomorphism 
\be
{}'x^\mu = e^{-\omega}x^\mu,\qquad {}'g_{\mu\nu}({}'x)=  e^{2\omega}g_{\mu\nu}(x)\qquad\hbox{and}\qquad {}'\Phi({}'x) = e^{r\omega}\Phi(x).\label{sd}
\ee

For a local transformation, parametrized by $\omega(x)$,  we then have to second order in $\omega$ 
\be
\de_\omega S[\Phi]=\int \left(\pa_\mu\omega\; V^\mu + \pa_\mu\omega\pa_\nu\omega\;L^{\mu\nu}\right)d^4x,\label{ol}
\ee
where $L^{\mu\nu}$ is a local, symmetric tensor field. For classical actions that are at most quadratic in the first derivative of the fields and that contain no higher derivative terms,  the r.h.s. of (\ref{ol}) is complete. In that sense $V_\mu$ and  $L^{\mu\nu}$ are the minimal set of currents present in a scale invariant theory. Generically, Lagrangians that contain higher powers of first derivatives and/or higher derivatives lead to non-unitarity theories although counter example exist (see e.g.  \cite{Smilga:2005gb,Rubakov:2013kaa} for discussions of such issues). 

A scale invariant theory is conformal if and only if  \cite{Coleman:1970je}
\be
V_\mu=\partial_\nu L_\mu^\nu\,.\label{VL}
\ee
This is what we would like to show. To this end  we will gauge this (Weyl-) scale symmetry \cite{Iorio:1996ad,Luty:2012ww,Skenderis}. For this we should require that the virial current is invariant under any additional gauge invariance of the action. Indeed, if the virial current is not gauge invariant, $\delta_\lambda V_\mu\neq 0$,  then $(\delta_\lambda  \delta_\sigma-    \delta_\sigma\delta_\lambda)S \neq 0$ so that we cannot simultaneously retain gauge- and local scale invariance\footnote{One may wonder if it is possible to nevertheless have conformal invariance, even if $V_\mu$ is not invariant under local gauge transformations. This means that there exist a local scalar $L$ such that $V_\mu=\partial_\mu L$ with $\delta_\lambda L={\cal{L}}\lambda + {\cal{L}}^\nu\partial_\nu\lambda$. It is not hard to see that this is incompatible with the existence of a gauge-invariant scale charge.}. Weyl gauging of the classical theory is then achieved through \cite{Iorio:1996ad} $\nabla_\mu\to \nabla_\mu + \Lambda^{\nu}_{\mu}W_\nu$, where $\nabla_\mu$ is the diffeomorphism covariant derivative. With this enlarged set of fields, $W_\mu$ is a source for the virial curent while $g_{\mu\nu}$ sources the energy-momentum tensor.  

We now turn to the quantum theory. Let ${\cal{O}}$ be a (collection of) operator(s) with well defined scale dimension, then
\be
\delta_\omega<{\cal{O}}>&=&i\int \omega(x) <T(x){\cal{O}}> d^4x\\
&=& <\delta_\omega{\cal{O}}>-i\int \partial_{\mu}\omega(x) <V^\mu(x){\cal{O}}> d^4x\,.
\ee
Dimensional arguments, Poincar\'e invariance and scale invariance lead to the general expression\footnote{There are five linearly independent structures for the $2$-point function which are reduced to two using symmetry and conservation.} for the 2-point function of $T_{\mu\nu}$,
\be
\<T_{\mu\nu} (p) T_{\alpha\beta}(q)\>&=&a(p,q)(p_\mu p_\nu-\eta_{\mu\nu}p^2)(p_\alpha p_\beta-\eta_{\alpha\beta}p^2)\\
&&\qquad+b(p,q)\frac{1}{2}\left[(p_\mu p_\alpha-\eta_{\mu\alpha}p^2)(p_\nu p_\beta-\eta_{\nu\beta}p^2)+\alpha\leftrightarrow\beta\right]\non\,,
\ee
where $a(p,q)=\delta^4(p+q)a(\frac{p^2}{\mu^2})$ and $b(p,q)=\delta^4(p+q)b(\frac{p^2}{\mu^2})$. 
Thus
\be
\<T (p) T(q)\>=(9a+4b)p^4\delta^4(p+q)\,.
\ee
If $a(x)$ and $b(x)$ are constants then $T(x)\equiv 0$ as a local operator and the theory is conformal. This is so because the 2-point function of $T$ reduces to a contact term that can be removed by a local counter term. On the other hand, for a unitary QFT  $<{\cal{O}}(x) {\cal{O}}(0)>\equiv 0$, implies ${\cal{O}}(x)\equiv 0$  (see eg. \cite{Reeh,Dymarsky:2013pqa,Nakayama:2013is} for more details). 
Conversely, if $T(x)$ does not vanish then, the correlator $<T(x)T(0)>$ diverges  like $1/|x|^8$ for small $|x|$. Thus, coupling to the metric $g_{\mu\nu}$ requires a counter term 
\be
S_{ct}\propto\frac{1}{\e}\int d^{4-\epsilon}x\;R^2\mu^{-\e}\,.\label{ctr}
\ee
Whence, $a(\frac{p^2}{\mu^2})=a_0+a_1\ln(p^2/\mu^2)$, with $a_1$ given by the residue of  $<{\cal{O}}(x) {\cal{O}}(0)>$ (see  \cite{Grinstein:2008qk,Skenderis} for details). 
At linear order we can gauge the Weyl-scale invariance through a coupling $\int W_\mu V^\mu$. Unless $V_\mu$ is invariant under all (gauge) symmetries other than the scale symmetry of the QFT in question, some of these symmetries will be broken explicitly by the Weyl-gauging. We therefore assume invariance of $V_\mu$ in what follows.  In order to see what sorts of counter terms may arise from this gauging beyond linear order we note that the operator product expansion of  $V_\mu(x)$ takes the form 
\be
V_\mu(x)V_\nu(0)\simeq a\frac{g_{\mu\nu}}{|x|^6}+b\left(\frac{g_{\mu\nu}}{|x|^6}-2\frac{{x_\mu x_\nu}}{|x|^8}\right)+\frac{L_{\mu\nu}(0)}{|x|^4}+\cdots
\ee
where $L_{\mu\nu}$, if present, is a local operator of dimension $2$.\footnote{Note that the only purpose of the OPE here is to see what local operators may arise as counter terms due to short distance singularities. In particular we are not assuming that a local dimension 2 operator $L_{\mu\nu}$ appears in the expansion (although we shall see shortly that it does). Operators of dimension less than two, while possibly present, do not lead to further counter terms and are therefore irrelevant for the present discussion.}  The terms proportional to the coefficients $a$ and $b$ give rise to counter terms
\be
S_{ct}\propto\frac{1}{\e}\int d^{4-\epsilon}x\;(a(\nabla_\alpha W^\alpha)^2+\frac{b}{2}F(W)_{\alpha\beta}F(W)^{\alpha\beta})\mu^{-\e}\,,\label{ctdw}
\ee
where $F(W)_{\alpha\beta}=\partial_\alpha W_\beta-\partial_\beta W_\alpha$ is the field strength of $W$ (which we might as well set to zero for the present discussion). If $L_{\mu\nu}$ does not vanish then there is furthermore a counter term
\be
S_{ct}\propto\frac{1}{\e}\int d^{4-\epsilon}x\;W^\alpha W^\beta L_{\alpha\beta}\;\mu^{-\e}\,.
\ee
Let us first focus on the scalar part, $L=\frac{1}{4} g^{\mu\nu}L_{\mu\nu}$. If $L$ does not vanish it gives rise to another counter term because 
\be 
< L(x)  L(0)>\propto \frac{ 1}{|x|^4} +\cdots
\ee
leading to the counter term 
\be
S_{ct}\propto\frac{1}{\e}\int d^{4-\epsilon}x\left( W_\alpha W^\alpha\right)^2\mu^{-\e}\,.
\ee
The traceless part of $L_{\mu\nu}$ cannot produce a local counter term\footnote{ This can be seen as follows: The general form of the 2-point function of the traceless part of $L_{\mu\nu}$ in momentum space is given by 
\be
<L_{\mu\nu}(p)L_{\alpha\beta}(-p)>&=&a \frac{p_\mu p_\nu p_\alpha p_\beta}{p^4}+b(g_{\mu\alpha}g_{\nu\beta}+g_{\mu\beta}g_{\nu\alpha}-\frac{1}{2}g_{\mu\nu}g_{\alpha\beta})\non\\
&&c(\frac{p_\beta p_\nu g_{\mu\alpha}}{p^2}+\frac{p_\alpha p_\nu g_{\mu\beta}}{p^2}+\frac{p_\beta p_\mu g_{\nu\alpha}}{p^2}+\frac{p_\alpha p_\mu g_{\nu\beta}}{p^2}-\frac{p_\beta p_\alpha g_{\mu\nu}}{p^2}-\frac{p_\mu p_\nu g_{\alpha\beta}}{p^2})\non
\ee
If only $b$ was non-vanishing and $L_{\mu\nu}$ of dimension 2, then this 2-point function could give rise to a local counter term. However, positivity of the imaginary part of the forward scattering amplitude implies that $b$ and $c$ are not independent (see \cite{Grinstein:2008qk,Dymarsky:2013pqa} for details). This in turn, is incompatible with a local counter term. }. From (\ref{ctr}) and(\ref{ctdw}) it follows that if the virial current does not vanish then the Weyl-gauge symmetry will be anomalous (see also \cite{Osborn:1991gm}). The general form of this anomaly is fixed by  the WZ-consistency condition \cite{Osborn:1991gm,Luty:2012ww}
\be
S_{WZ}[ \delta_\omega g_{\mu\nu},W_\mu-\partial_\mu \omega]\propto \int \sqrt{g}\,\omega \left(\frac{1}{6}R+\nabla_\alpha W^\alpha- W_\alpha W^\alpha\right)^2 d^4 x\,,
\ee
which, in turn, fixes the relative coefficients of the various counter terms, ie. \cite{Skenderis}
\be
S_{ct}\propto\frac{1}{\e}\int d^{4-\epsilon}x\left(\frac{1}{6}R+\nabla_\alpha W^\alpha- W_\alpha W^\alpha\right)^2\mu^{-\e}\label{ctf}\,.
\ee
In particular, for non-vanishing virial current  eqn (\ref{ctf}) then implies the existence of a local dimension $2$ hermitean operator, $L_{\mu\nu}=g_{\mu\nu}L$, in agreement with our classical considerations at the beginning. The logic underlying this conclusion is that the presence of a counter term implies the existence of a local operator that gives rise to it. Furthermore, we conclude from (\ref{ctf}) that the coefficients of the counter terms arising from the singularity of  the two-point functions of $V_\mu$ and $L$ and therefore the corresponding residues are not independent, 
\be
<L(p)L(q)>={\cal{A}}(p,q)\quad\hbox{and}\quad <V_\mu (p) V_\nu(q)>= {\cal{A}}(p,q)p_\mu p_\nu+{\cal{B}}(p,q)(g_{\mu\nu}p^2 -p_\mu p_\nu)\,,
\ee
where  ${\cal{A}}(p,q)=a\delta^4(p+q)\log(\frac{p^2}{\mu^2})$ and ${\cal{B}}(p,q)=b\delta^4(p+q)\log(\frac{p^2}{\mu^2})$. To continue we then define ${\cal{O}}_\mu(p)\equiv V_\mu (p)-(\partial_\mu L)(p)$ for which we have from (\ref{ctf}) modulo contect terms
\be
<{\cal{O}}_\mu (p) {\cal{O}}_\nu(q)>= {\cal{B}}(p,q)(g_{\mu\nu}p^2 -p_\mu p_\nu)\,.
\ee
This implies that for a unitary theory the longitudinal part of ${\cal{O}}_\mu$ vanishes as an operator or,  equivalently, that  $V_\mu=\partial_\mu L$ up to a transverse contribution. Consequently, we have for the trace of the Belinfante stress tensor, $T(x)=\nabla^2 L(x)$ which is just the condition for conformal invariance. 
 
Let us now re-examine the assumptions made to obtain this result. The starting point was a scale-invariant QFT that can be coupled minimally to the metric tensor. So, we assume a conserved, symmetric stress tensor. Then, assuming diffeomorphism invariance,  we reformulated scale-invariance as the invariance under  $g_{\mu\nu}\to e^{2\omega}g_{\mu\nu}$, $\Phi \to e^{-\Delta_\Phi\omega}\Phi$ . Next, we gauged this symmetry by coupling an external Weyl gauge potential, $W_\mu$ to the virial current $V_\mu$. We note in passing that  $V_\mu$ is not conserved since scale invariance requires the transformation of the external metric field as well. For the Weyl-gauging to be consistent with other possible symmetries of the theory we need to assume that the virial current is invariant under the corresponding transformations. 
An illuminating example is that of a 2-form potential $A_{\mu\nu}$ where the virial current $V_\mu\propto A_{\alpha\beta}F^{\alpha\beta}_\mu$ can be written as a derivative only in the gauge $\partial^\alpha A_{\alpha\beta}=0$. However, assuming this gauge condition $A_{0 i}$ becomes a non-local function of $A_{i j}$, $i,j=1,2,3$. Furthermore, one finds that while  energy, momentum and dilatation charge are gauge-invariant, the charges generating special conformal transformations are not. We note in passing that $F_{\alpha\beta\mu}$ nevertheless has a well defined transformation under conformal transformations since it is dual to a free field, $F_{\alpha\beta\mu}=\epsilon_{\alpha\beta\mu}^{\;\;\;\;\;\;\;\;\nu}\partial_\nu\phi$. It transforms as a descendant rather than a primary.

 \bigskip
\noindent{\bf Acknowledgements:}\\
\smallskip
I would like to thank S. Konopka, Z. Komargodski, D. Ponomarev, A. Schwimmer, S. Theisen and A. Vikmann  for helpful discussions. This work  was supported by the DFG Transregional Collaborative Research Centre TRR 33, the DFG cluster of  excellence "Origin and Structure of the Universe". 

\end{document}